# Manipulating single excess electrons in monolayer transition metal dihalide

Min Cai[1], Yunfan Liang[2], Zeyu Jiang[2], Mao-Peng Miao[1], Zhen-Yu Liu[1], Wen-Hao Zhang[1], Xin Liao[1], Wei Cheng[1], Damien West[2], Shengbai Zhang[2], and Ying-Shuang Fu[1*]

1. School of Physics and Wuhan National High Magnetic Field Center, Huazhong University of Science and Technology, Wuhan 430074, China

2. Department of Physics, Applied Physics and Astronomy, Rensselaer Polytechnic Institute, Troy, NY, 12180, USA

*yfu@hust.edu.cn

**Polarons are entities of excess electrons dressed with local response of lattices, whose atomic-scale characterization is essential for understanding the many body physics arising from the electron-lattice entanglement, but yet difficult to achieve. Here, using scanning tunneling microscopy and spectroscopy (STM/STS), we show the visualization and manipulation of single polarons with different origin, i.e., electronic and conventional polarons, in monolayer $CoCl_2$, that are grown on HOPG substrate via molecular beam epitaxy. Four types of polarons are identified, all inducing upward local band bending, but exhibiting distinct appearances, lattice occupations, polaronic states and local lattice distortions. First principles calculations unveil three types of polarons are stabilized by electron-electron interaction. The type-4 polaron, however, are driven by conventional lattice distortions. All the four types of polarons can be created, moved, erased, and moreover interconverted individually by the STM tip, allowing precise control of single polarons unprecedently. This finding identifies the rich category of polarons**

**and their feasibility of manipulation in CoCl$_2$, which can be generalized to other transition metal halides.**

The interaction of electrons and lattice plays a pivotal role in the many body physics of solids [1,2]. While the lattice vibrations can be treated perturbatively to the electron transport in the weak-coupling limit, the lattice anti-adiabatically responds to the excess electron in the strong coupling limit [3]. In a dielectric lattice, strong electron-lattice coupling results in the displacement of ions adjacent to the excess electron, creating a potential well to get the electron trapped [4,5]. As a result, the excess electron is dressed by local lattice distortions when propagating through polarizable solids, forming a composite quasiparticle, named as polaron [6-9]. Apart from the conventional formation mechanism of ion-displacement, the trapping potential could be triggered dominantly via electron-electron interaction. In such a case, the presence of the additional electron induces electric dipole pointing toward it without the need of significant ion displacement [10], which lowers its kinetic energy and forming a distinct type of electronic polaron [11].

Polarons have attracted extensive interests in multidiscipline fields [8,9], because of their key impacts on many physical processes, as exemplified in charge transport [12], colossal magnetoresistance [13], high temperature superconductivity [14], as well as on chemical reactivity [15,16] and functional materials with thermoelectric and multiferroic behavior [17,18]. Depending on their multi-facet properties, polarons can be detected with various probes, manifesting themselves as large mass in charge transport [19], associated band dispersion in angle-resolved photo-emission spectra [20-22], absorption and luminescence spectra of their transition between the ground state and excited state [23,24], as well as their spin character in electron spin resonance spectra [25,26], etc. Despite of their existence probed with above ensemble averaged techniques, a direct visualization and spectroscopic characterization of individual polarons at atomic-scale is highly desirable for elucidating their properties [27,28]. This is particular important in material systems where different types of polarons coexist, exhibiting distinct atomistic and electronic structures [29].

STM has the capability of atomic resolution imaging and spectroscopic characterization of the local density of states, which is highly suited for the probe of single polarons. However, the dielectric crystals hosting polarons are insulators, posing challenge for STM tunneling measurement. This restricts pioneering STM studies on polarons to systems of narrow gap semiconductors or doped insulators [29-31]. To overcome that hurdle, an alternative arena is to study polaron thin films on conductive substrate, which allows electron tunneling through [32]. However, the conductive substrate may enhance charge screening and strongly interacts with the thin films, potentially destabilizing the polarons. Our strategy is to use graphene or HOPG substrate, which have not only low carrier density but also weak van der Waals (vdW) interaction with the supporting films [33]. The system of choice is monolayer transition metal dihalide, which possesses ionic bonding and strong correlation effects [34]. Both factors constitute the essential ingredients for the formation of conventional polarons driven by electron-phonon coupling and electronic polaron originating from electron-electron coupling.

In this work, we report the visualization and manipulation of single polarons in monolayer $CoCl_2$ on HOPG substrate with STM. Four types of polarons are discovered in $CoCl_2$, where two types are centered on Co sites with different apparent heights and hopping barriers, and the other two types are centered on top-Cl sites, but with different appearances and polaronic states. One of the Cl-centered polaron induces evident lattice distortion that extends to several lattice units, and is ascribed as conventional polaron. In contrast, the other three types, as unveiled with first principles calculations, are electronic polarons. More importantly, all types of polarons can be manipulated, with controlled creation, movement, and annihilation. Interconversion among different types of polarons is also achieved, demonstrating versatile tunability of this polaron system.

The experiments were performed with a custom-made cryogenic Unisoku 1500 STM system [35]. Monolayer $CoCl_2$ films are grown by MBE on HOPG substrate. The first principles calculation is carried out with hybrid functional HSE06 approach. Detailed descriptions of the experiments and the calculations are depicted in Methods.

Bulk $CoCl_2$ is a van der Waals crystal that belongs to trigonal $\bar{R}3m$ space group [36]. Its each vdW layer consists of a triangular lattice of Co layer sandwiched between two Cl layers, with each Co cation octahedrally coordinated by six Cl anions, forming 1-T structure [Figure 1(a)]. Figure 1(b) shows a typical topographic image of monolayer $CoCl_2$ film on HOPG substrate. The apparent height of the monolayer film is measured as 780 pm at the imaging bias of 2 V and indicate prominent bias dependence (Fig. S1), demonstrating its distinct spectroscopic features from the substrate. Atomic resolution image of $CoCl_2$ imaged at 0.3 V displays a triangular lattice of Cl atoms, whose measured lattice constant 350 pm is consistent with the reported value of 354 pm [Figure 1(c)] [36]. There is a moiré pattern with a 3×3 periodicity superimposed on the atomic lattice, as is also seen from its fast Fourier transformation image [Figure 1(c), inset]. Multiple moiré patterns are observed upon changing the relative angle between $CoCl_2$ and graphene layers [Fig. S2].

When imaging the monolayer $CoCl_2$ at a larger bias of 1.2 V, there appear two types of depressed defect-like entities with different apparent depth coexisting with another type of protruded entities randomly distributed over the film [Figure 1(d)]. As will be shown later, those three types of defect-like entities are all polarons, instead of crystal defects. Hereafter, the shallow and deep entities are named as type-1 and type-2 polarons, respectively, and the protruded entities are type-3 polarons. The apparent depths of the type-1, type-2 and type-3 polarons are measured as - 1.1 Å, - 2.3 Å and 0.4 Å, respectively [Figure 1(e)]. Moreover, their appearances change drastically with imaging bias. For both the type-1 and type-2 polarons, their lateral size increases monotonically with decreasing bias from 1.2 V to 0.4 V, and becomes completely invisible with the bias setting at the band gap of $CoCl_2$ [Fig. S3]. The type-3 polaron appears as a protrusion at biases above 1 V and below -1.5 V, but becomes a depression and indistinguishable with the type-1 polaron between 0.5 V and 1 V [Figs. S4, S5].

Next, we characterize the electronic properties of the polarons. The spectrum of the bare monolayer $CoCl_2$ features a conduction band located at 0.25 eV and the valence band edge should be below the measured spectroscopic range of -3 eV [Fig. 3(f)]. Upon approaching the three types of polarons, the spectra all exhibit evident upward band bending, but with a different magnitude [Figs. 1(g) and S6]. Specifically, the conduction band edge shifts to 0.65 eV (1.3 eV) for the type-1 (type-2) polaron [Fig. S6]. Compared to the type-1 polaron, the type-3 polaron induces similar energy shift of conduction band edge, above which its conductance is larger [Fig. 1(g)]. The local band bending around the polarons causes depletion of density of states at energies corresponding to the conduction band of $CoCl_2$, which consequently renders the apparent STM image of the polarons as depressions. The type-2 polaron has a larger band bending, conforming to its deeper apparent depth, than that of the type-1 polaron. While the type-1 and type-2 polarons have no noticeable polaronic states, the type-3 polaron exhibits clear polaronic sates below - 1.5 eV with a peak at - 2 eV [Figs. 1(f) and (g)].

The upward local band bending implies the polarons host additional electrons, elevating the local chemical potential. It is noted that crystal defects of charge acceptors could also accommodate additional electrons. To discern the two scenarios, we perform tip manipulations to the defect-like entity. Upon the tip is on top of a type-1 entity and with the tunneling junction setting at 3.0 V and 7 nA, the tip height becomes unstable among different values [Fig. S7], implying state switching of the entity. Subsequent imaging the same field of view at low bias shows the type-1 entity is completely annihilated [Figs. 2(a),(b)]. This unambiguously excludes the defect-like entity as a real crystal defect, but rather suggests it as a polaron. The type-1 polaron can also be created by applying a high negative bias of - 3.5 eV for a duration of 5 seconds [Figs. 2(c),(d)]. The created polaron indicates identical morphology and spectroscopic features as the naturally formed ones. Distinct to crystal defects which can hardly be manipulated at low temperatures, polarons can easily hop among different sites. Indeed, we can laterally move the type-1 polaron, which could stably follow the tip trajectory by setting the junction at 3.0 V/3 nA with a moving speed of 50 pm/s. This allows the feasibility of tuning the local polaron density. As is demonstrated in Fig. S8, a clean region can be obtained by moving the polaron out of the area of interest with tip manipulation.

The type-2 polarons can also be manipulated with the tip, which however bear some differences compared to the type-1 polarons. First, the type-2 polaron has a lower hopping barrier. As is seen from Fig. S7, the type-2 polarons are laterally manipulated even with the scanning condition of 2.5 V/10 pA, while the type-1 polarons in the same field of view are all kept stationary. Second, in contrast to the type-1 polaron, the lateral manipulation of the type-2 polarons can be performed with both bias polarities, but showing different behaviors. Specifically, a positive bias (for instance, 4 V) causes many surrounding type-2 polarons within a large spatial range to simultaneously accumulate toward the tip location. And, a negative bias of -2 V solely moves a single type-2 polaron without disturbing neighboring polarons. These characters allow it feasible to move the type-2 polarons both effectively and precisely with different bias polarities. Third, type-2 polarons cannot be directly created or annihilated, but have to go through the type-1 polaron states, as depicted below.

We find that the different types of polarons can interestingly be converted in a reversible manner. Fig. 2(g) shows that the field of view are dominated with type-1 polarons. By positioning the tip at a type-1 polaron at 3.0 V/4 nA for a prolonged time, a current jump occurs, after which it converts to a type-2 polaron [Fig. 2(h)]. With that manipulation protocol, we could reproducibly perform the transition on another type-1 polaron, as is seen in Fig. 2(i). The reversed transition from type-2 polarons to type-1 polarons can be realized with a different tunneling parameter of - 2.5 V/10 pA, as is indicated in Figs. 2(g-i) with reversed black arrows. Moreover, the type-3 polaron can be transformed into a type-2 polaron by tip manipulation [Fig. S9].

Besides the above three types of polarons, a new polaron species, designated as type-4 polaron, are observed. The type-4 polarons are generated via the type-1 polarons during tip manipulations, instead of being naturally formed during growth. Fig. 3(a) shows three type-1 polarons, one of which abruptly changes when imaged with a high negative bias of -4.5 V at 77 K and is dragged by the scanning tip subsequently [Fig. 3(b)]. Imaging the same area again at low bias indicates the type-1 polaron has converted to a type-4 polaron, which exhibits a triangular-shape, in contrast to the rounded shape of the type-1 polaron. We find such polaron conversion never happens at 5 K, whose transition probability, however, increases with elevating the measurement temperature. At 200 K, the type-4 polarons can be generated not only from the type-1 polarons with a reduced bias of -3.5 V, but also directly out of the $CoCl_2$ films during scanning with -5 V [Fig. S10]. Apart from their creation and movement, the type-4 polarons can also be erased via tip manipulation [Fig. S11]. The type-4 polaron is also an electron polaron, because it induces similar upward local band bending [Figure 3(d)]. There appears an obvious polaronic peak at -1.18 eV [Fig. 3(e)]. Its STM images change prominently with bias, which appear as bright protrusions at the energies of the polaronic states and change to depress spots at other energies [Fig. S12].

We then carry out high resolution STM imaging to reveal the atomistic structures of the different types of polarons. To determine the occupation site of the polarons, atomic resolution of $CoCl_2$ should be obtained simultaneously, which requires low bias STM imaging. However, lateral sizes of the type-1 and type-2 polarons significantly enlarge with decreasing bias, and become invisible at low bias. Nevertheless, after decorating the tip apex with controlled dipping onto $CoCl_2$, atomic resolution of the type-1 and type-2 polarons can be imaged and display similar features. Namely, three Cl atoms surrounding centers of both the type-1 and type-2 polarons become bright, forming a trimmer geometry. The trimmers are all oriented along the same direction of the Cl lattice, demonstrating their identical occupation sites, presumably on Co-sites. Moreover, no lattice distortions are discernable around the type-1 and type-2 polarons.

Atomic resolutions of the type-3 and type-4 polarons, imaged with a normal W tip, indicate they are both centered on Cl-sites, but with different appearances. While the type-3 polaron appear as a single atomic protrusion on Cl, the type-4 polaron features a dark center neighboring 6 bright Cl atoms. Furthermore, despite that the Cl lattice surrounding the type-3 polaron barely distorts, the type-4 polaron induces evident local lattice distortion, whose radial extension is up to 5 lattice units.

To reveal the nature of different types of polarons, we perform first-principles calculations with hybrid functional HSE06 approach on single electron doped $CoCl_2$ monolayer. Two different types of polarons are obtained with their simulated STM images shown in Fig. 5(a) and (b), which are in good agreement with Fig. 4. We find that both type-1 and type-2 polarons in experiments are associated with Fig. 5(a) which manifests as a Co-centered triangle, while Fig. 5(b) is consistent with the type-3 polaron which is a Cl-centered bright point. From the charge distribution of individual polaronic state shown in Fig. 5(c-f), we find that the bright spots in STM of Co- and Cl-centered polaron come from the first neighbor and the central Cl atoms in the top layer, respectively. According to the side view, there are rather limited electronic population on the first neighbor Cl atoms for the Co-centered polaron, thus the polaronic states in the experimental STS might be too small and easily hidden by the background noise. Instead, for the Cl-centered polaron, the central Cl atom holds large electron density, so the polaronic state is strong enough to be observed in STS. Interestingly, even that the geometry center of the type-3 polaron is on Cl atom, it's actually constructed from 3d orbitals of three Co atoms which are first neighbor to the central Cl atom, and the occupation on the central Cl atom is rather limited. This is consistent with the calculated local density of states that the in-gap peaks are dominated by Co-3d orbitals, as shown in Fig. 5(g) and (h). The polaron formation energy, which is defined as the energy difference relative to the delocalized state, is calculated to be -217 and -292 meV, respectively. The large negative values and as well as the similar magnitude of polaron formation energies explains why they can co-exist stably within a considerable range of temperature.

Furthermore, we find that Co- and Cl-entered polarons are both electronic polarons, which can form spontaneously only with electron-electron interaction in an undistorted lattice, i.e., when the structural translational symmetry is preserved. By removing the lattice distortions in our calculation, the polaron formation energy of electronic polaron is calculated to be -25 and -53 meV, indicating that these polarons can be created even without phonon response. To some extent, this may explain why both types of polarons can be obtained naturally within the growth of sample. While the type-4 polarons are not yet captured by calculations, their induced large lattice distortions suggest their conventional origin of electron-phonon coupling. They are not formed naturally, presumably due to the inadequate response time of lattice vibrations to the injected electrons. However, those type-4 polarons are transformed from the type-1 polaron, implying the electronic polaron increases the dwelling time of the excess charge in the $CoCl_2$ lattice, and acting as a precursor state to form the conventional polaron.

Finally, we briefly discuss the difference between type-1 and type-2 polarons. In the DFT result, there is only one Co-centered polaron. However, the appearance of the substrate can break the inversion symmetry of the polaron and cause different types of polaronic structures. Such interaction probably changes the polaronic states and the STM images. Another possibility is that they have different spin polarization. The polaron with different polarization can have different electronic structure and ends up in different STM image. However, our measurement suggests the type-1 and type-2 polarons persist up to 200 K, which should above the Curie temperature of $CoCl_2$. This makes the later scenario unlikely.

In summary, we have visualized and manipulated individual polarons in monolayer $CoCl_2$. Four types of polarons with different origin, i.e. electronic polaron and conventional polaron, are identified from their distinct signatures of appearances, polaronic states, occupation cites and local lattice distortion. Those different types of polarons manifest stunning versatility of manipulation and interconversion. The rich category of polarons envision to exist in other transition metal dihalide films, since they share similar electronic and structural characteristics. As such, we have examined monolayer $FeCl_2$ (Fig. S13), and indeed observed similar polarons. Moreover, the manipulation of the excess charges trapped inside the polarons render the feasibility of locally tuning the charge density of the films, which acts as a unique knob for controlling other many body states, such as exciton condensation [37]. In addition, the spin properties of the polarons and their exchange interaction with the magnetism of the monolayer insulator constitute another interesting subject of future investigations.

**Methods**

The monolayer $CoCl_2$ film is grown on HOPG. The HOPG substrate was cleaved *ex situ* and further degassed in a vacuum chamber at approximately 1170 K for 0.5–2 h before growth. High purity ultra-dry $CoCl_2$ powder (99.99% Alfa Aeser) is evaporated at 523 K from a home-made k-cell evaporator and the substrate temperature is kept around 470 K during the sample growth. The base pressure is better than $5 \times 10^{-9}$ torr. The STM measurement is conducted at 77 K if not stated specifically. An electrochemically etched W wire was used as the STM tip, which had been cleaned on a Ag(111) surface prior to conducting the measurements. The STS is taken by the lock-in technique with a modulation of 21.2 mV (rms) at 983 Hz.

The VASP code [38] is applied to perform the density functional theory (DFT) calculations with Projector augmented wave (PAW) potentials. To describe the transition metal system with d orbital more physically, all the calculation is done with the Hybrid functional (HSE06) [39]. The polarons are simulated in a 5×5×1 supercell with one extra electron. The kinetic energy cut off is 400 eV and the total energy is converged to $10^{-5}$ eV in self-consistent calculations. During structure relaxation, the lattice constant is fixed, while all atoms are free to move. The convergence condition of the force is set to be -0.02 eV/ Å.

**Acknowledgement:** This work was funded by the National Natural Science Foundation of China (Grant Nos. U20A6002, 11874161, 12047508), the National Key Research and Development Program of China (Grant Nos. 2017YFA0403501, and 2018YFA0307000), the U.S. Department of Energy, Office of Science, Office of Basic Energy Sciences under Award Number DESC-0002623. The supercomputer time sponsored by the National Energy Research Scientific Center (NERSC) under DOE Contract No. DE-AC02-05CH11231 and the Center for Computational Innovations (CCI) at RPI are also acknowledged.

**Note added:**

During the preparation of this manuscript, we were aware of a related study posted on arXiv:2205.10731, where the details of the observations and data interpretations are different from ours.

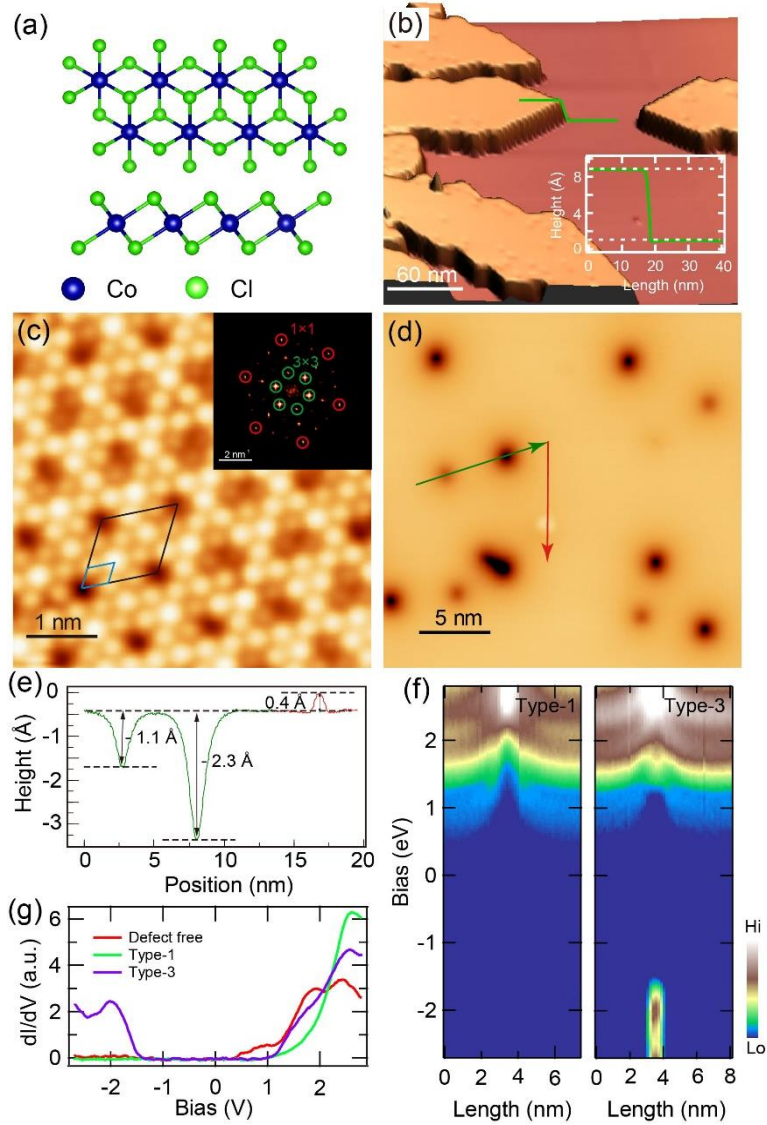

Figure 1 Topography and spectra of polarons. (a) Top and side view of the crystal structure of monolayer $CoCl_2$. (b) Large-scale STM topographic image of $CoCl_2$ film ($V_s$ = 2 V, $I_t$ = 5 pA). The inset shows a line profile taken along the green line. (c) Atomic resolution of $CoCl_2$ ($V_s$ = 0.3 V, $I_t$ = 20 pA). The blue and black rhombus indicate the unit lattice of the Cl atoms and a 3 × 3 moiré pattern, whose fast Fourier transform is shown in the inset. (d) STM image of three types of polarons ($V_s$ = 1.2 V, $I_t$ = 10 pA). (e) Line profile of the three types of polarons taken along the green and red arrows marked in (d). (f) 2D conductance plot across the type-1 and type-3 polarons. Spectroscopic condition: $V_s$ = 2.8 V, $I_t$ = 150 pA. (g) Point spectra measured on the type-1, type-3 polaron center and the polaron-free area of the film.

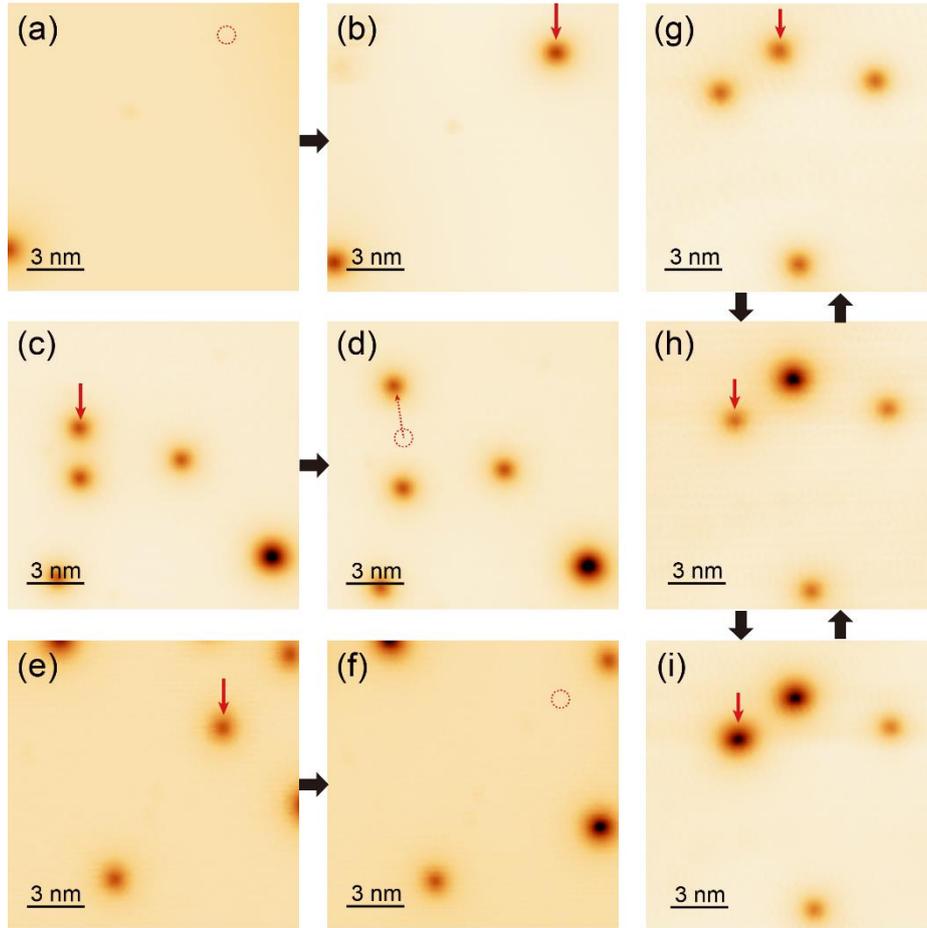

Figure 2 Manipulation and interconversion of polarons. (a, b) STM images taken before and after creating a type-1 polaron. (c, d) STM images taken before and after moving a type-1 polaron. (e, f) STM images taken before and after erasing a type-1 polaron. (g-i) Sequence of STM images showing interconversion between type-1 and type-2 polarons. Image conditions for all images: $V_s$ = 1.2 V, $I_t$ = 10 pA. The red dashed circles mark the original location of the polarons before manipulation. The red arrows mark the manipulated polarons.

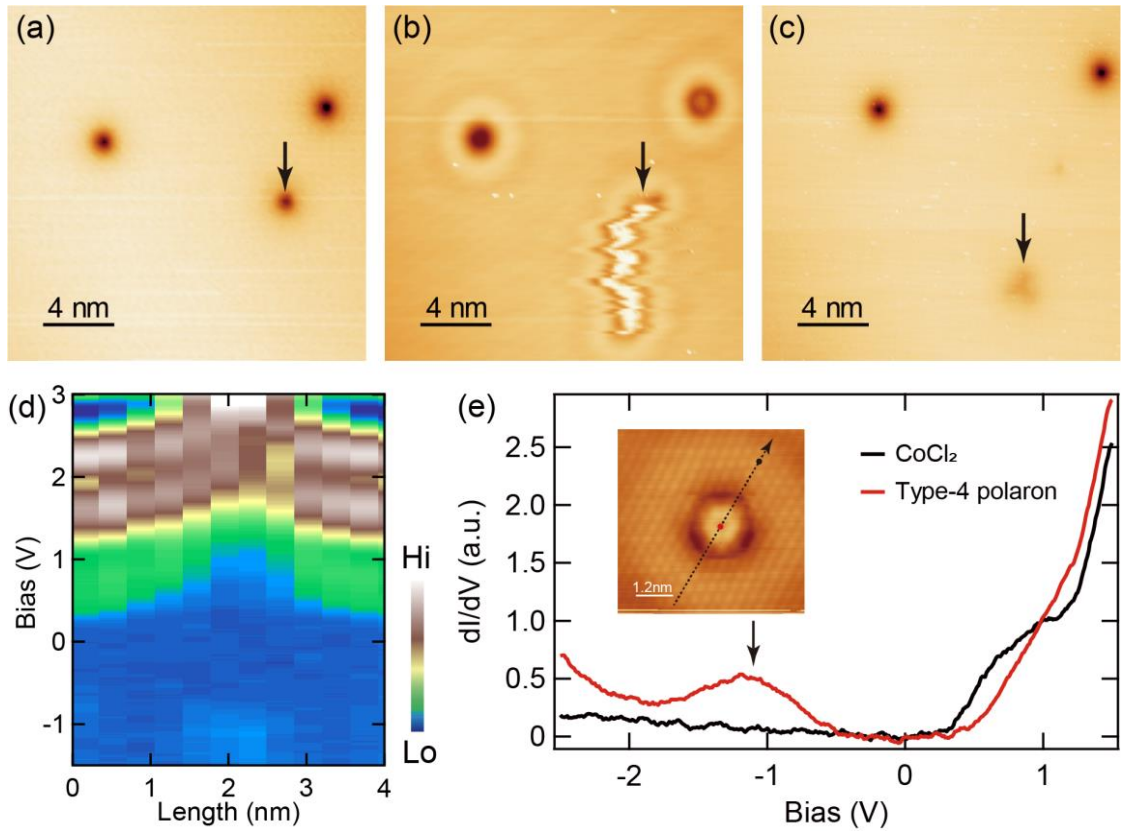

Figure 3 Type-4 polaron. (a) STM image of three type-1 polaron. (b) STM image taken at $V_s = -4.5$ V, $I_t = 100$ pA, showing conversion of a type-1 polaron to type-4 polaron. (c) STM image taken after the generation of the type-4 polaron. Image conditions for (a, c): $V_s = 1$ V, $I_t = 10$ pA. (d) 2D conductance plot taken along the black dashed arrow in the inset of (e). (e) Point Spectra taken at defect free area and the type-4 polaron center. The black arrow marks the its polaronic state.

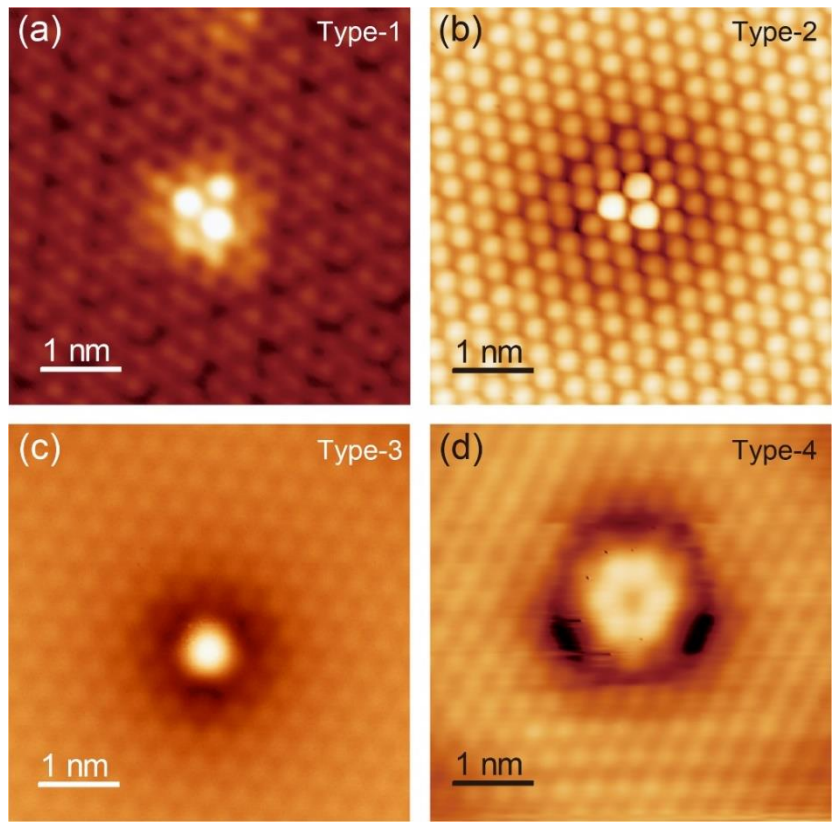

Figure 4 Atomic resolution of polarons. (a) Atomic resolution ($V_s = -1$ V, $I_t = 100$ pA) of type-1 polaron. (b) Atomic resolution ($V_s = -0.2$ V, $I_t = 50$ pA) of type-2 polaron. (c) Atomic resolution ($V_s = 1.4$ V, $I_t = 10$ pA) of type-3 polaron. (d) Atomic resolution ($V_s = -2.2$ V, $I_t = 100$ pA) of type-4 polaron.

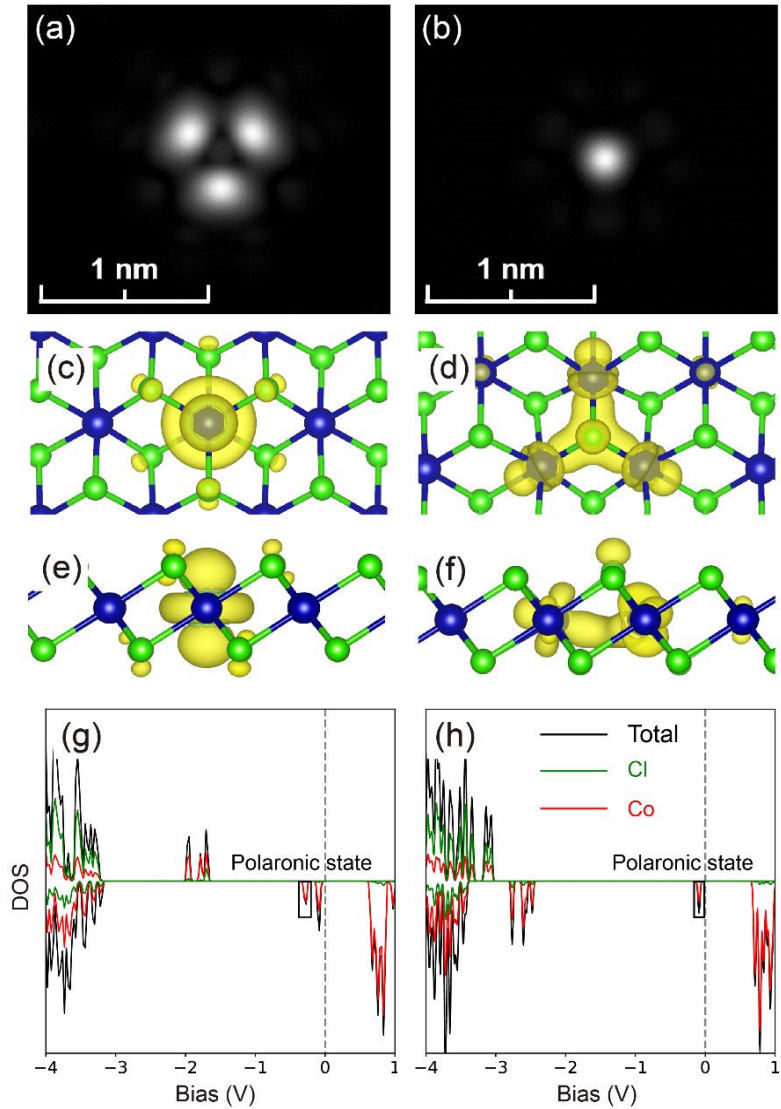

Figure 5 Calculated polarons. (a,b) Simulated STM images for Co (a) and Cl (b) centered polaron. (c,d) Top view of the charge distribution of polaronic state for Co (c) and Cl (d) centered polaron. (e,f) Side view of the charge distribution of polaronic state for Co (e) and Cl (f) centered polaron. (g,h) Total and partial DOS for Co (g) and Cl (h) centered polaron.

**References:**

[1] Snoke, D. W. Solid State Physics: Essential Concepts. *Addison-Wesley* (2009).

[2] Giustino, F. Electron-phonon interactions from first principles. *Rev. Mod. Phys.* **89**, 015003 (2017).

[3] Fröhlich, H. Electrons in lattice fields. *Adv. Phys*. **3**, 325-361 (1954).

[4] Landau, L. D. On the Motion of Electrons in a Crystal Lattice. *Phys. Z. Sowjetunion* **3**, 664 (1933).

[5] Pekar, S. I. Local quantum states of electrons in an ideal ion crystal. *Zh. Eksp. Teor. Fiz.* **16**, 341–348 (1946).

[6] Holstein, T. Studies of polaron motion: Part II. The "small" polaron. *Ann. Phys.* **8**, 343 (1959).

[7] Devreese, J. T. Polarons. *Encyclopedia of Applied Physics* **14**, 383-409 (1996).

[8] Franchini, C., Reticcioli, M., Setvin, M. *et al*. Polarons in materials. *Nat. Rev. Mater.* **6**, 560–586 (2021).

[9] D. Emin, Polarons (Cambridge University Press, 2013).

[10] Berciu, I. Elfimov & G. A. Sawatzky Electronic polarons and bipolarons in iron-based superconductors: The role of anions. *Physical Review B* **79**, 214507 (2009).

[11] Liang, Y. *et al*. Electronic Polaron in Two-Dimensional Transition Metal Halides. Arxiv: 2110.01790.

[12] Coropceanu, V. *et al*. Charge transport in organic semiconductors. *Chem. Rev.* **107**, 926–952 (2007).

[13] Teresa, J. M. D. *et al*. Evidence for magnetic polarons in the magnetoresistive

perovskites. *Nature* **386**, 256–259 (1997).

[14] A. S. Alexandrov & A. B. Krebs, Polarons in high temperature superconductors. *Soviet Physics Uspekhi*, **35**, 345 (1992).

[15] Yin, W.-J., Wen, B., Zhou, C., Selloni, A. & Liu, L.-M. Excess electrons in reduced rutile and anatase $TiO_2$. *Surf. Sci. Rep*. **73**, 58–82 (2018).

[16] Reticcioli, M. *et al*. Interplay between adsorbates and polarons: CO on rutile $TiO_2$(110). *Phys. Rev. Lett*. **122**, 016805 (2019).

[17] Wang, M. *et al*. Thermoelectric Seebeck effect in oxide-based resistive switching memory. *Nat. Commun*. **5**, 4598 (2014).

[18] Miyata, K. & Zhu, X.-Y. Ferroelectric large polarons. *Nat. Mater*. **17**, 379–381 (2018)

[19] Nagels, P., Denayer, M. & Devreese, J. Electrical properties of single crystals of uranium dioxide. *Solid State Commun*. **1**, 35–40 (1963).

[20] Moser, S. *et al*. Tunable polaronic conduction in anatase $TiO_2$. *Phys. Rev. Lett*. **110**, 196403 (2013).

[21] Strocov, V. N., Cancellieri, C. & Mishchenko, A. S. Electrons and Polarons at Oxide Interfaces Explored by Soft-X-Ray ARPES (Springer, 2018).

[22] M. Kang, S. W. Jung, W. J. Shin, Y. Sohn, S. H. Ryu, T. K. Kim, M. Hoesch & K. S. Kim. Holstein polaron in a valley-degenerate two-dimensional semiconductor. *Nat. Mater*. **17**, 676 (2018).

[23] Vura-Weis, J. *et al*. Femtosecond M2,3-edge spectroscopy of transition-metal oxides: photoinduced oxidation state change in α-$Fe_2O_3$. *J. Phys. Chem. Lett*.**4**,

3667–3671 (2013).

[24] Grenier, P., Bernier, G., Jandl, S., Salce, B. & Boatner, L. A. Fluorescence and ferroelectric microregions in KtaO$_3$. *J. Phys. Condens. Matter*. **1**, 2515–2520 (1989).

[25] Chiesa, M., Paganini, M. C., Livraghi, S. & Giamello, E. Charge trapping in TiO$_2$ polymorphs as seen by electron paramagnetic resonance spectroscopy. *Phys. Chem. Chem. Phys*. **15**, 9435–9447 (2013).

[26] Shengelaya, A., Zhao, G.-m, Keller, H. & Müller, K. A. EPR evidence of Jahn-Teller polaron formation in La$_{1-x}$Ca$_x$MnO$_{3+y}$. *Phys. Rev. Lett*. **77**, 5296–5299 (1996).

[27] Ronnow, H. M., Renner, C., Aeppli, G., Kimura, T. & Tokura, Y. Polarons and confinement of electronic motion to two dimensions in a layered manganite. *Nature* **440**, 1025-1028 (2006).

[28] Wu, L., Klie, R. F., Zhu, Y. & Jooss, C. Experimental confirmation of Zener-polaron-type charge and orbital ordering in Pr$_{1-x}$Ca$_x$MnO$_3$. *Phys. Rev. B* **76**, 174210 (2007).

[29] Setvin, M. *et al*. Direct view at excess electrons in TiO$_2$ rutile and anatase. *Phys. Rev. Lett*. **113**, 086402 (2014).

[30] Yim, C. M. *et al*. Engineering polarons at a metal oxide surface. *Phys. Rev. Lett*. **117**, 116402 (2016).

[31] Y. Mao, X. Ma, D. Wu, C. Lin, H. Shan, X. Wu, J. Zhao, A. Zhao, & B. Wang, Interfacial Polarons in van der Waals Heterojunction of Monolayer $SnSe_2$ on $SrTiO_3$ (001). *Nano Lett*. **20**, 8067 (2020).

[32] Repp, J. Meyer, G., Olsson, F.E., Persson, M. Controlling the Charge State of Individual Gold Adatoms. *Science* **305**, 493-495 (2004).

[33] X. Yang, J.-J. Xian, G. Li, N. Nagaosa, W.-H. Zhang, L. Qin, Z.-M. Zhang, J.-T. Lü, and Y.-S. Fu, Possible Phason-Polaron Effect on Purely One-Dimensional Charge Order of $Mo_6Se_6$ Nanowires, *Phys. Rev. X* **10**, 031061 (2020).

[34] L. Peng, J. Z. Zhao, M. Cai, G.-Y. Hua, Z.-Y. Liu, H.-N. Xia, Y. Yuan, W.-H. Zhang, G. Xu, L.-X. Zhao, Z.-W. Zhu, T. Xiang, and Y.-S. Fu, Mott phase in a van der Waals transition-metal halide at single-layer limit. *Phys. Rev. Research* **2**, 023264 (2020).

[35] Xian, J.J. et al., Spin mapping of intralayer antiferromagnetism and field-induced spin reorientation in monolayer $CrTe_2$, *Nat. Commun*. **13**, 257 (2022).

[36] Grimme, H.; Santos, J.A. The Structure and Colour of Anhydrous Cobalt Chloride, $CoCl_2$, at Room and very Low Temperatures. *Z. Kristallogr*. **88**, 136–141 (1934).

[37] Z. Y. Jiang, Y. C. Li, W. H. Duan & S. B. Zhang, Half-Excitonic Insulator: A Single-Spin Bose-Einstein Condensate. *Phys. Rev. Lett*. **122**, 236402 (2019).

[38] Kresse G, Furthmüller J. Efficient iterative schemes for ab initio total-energy calculations using a plane-wave basis set. *Physical review B* **54**, 11169 (1996).

[39] Heyd J, Scuseria G E, Ernzerhof M. Hybrid functionals based on a screened Coulomb potential. *The Journal of chemical physics* **118**, 8207-8215 (2003).